\pdfoutput=1
\documentclass{JINST}
\let\ifpdf\relax

\usepackage{epsfig}
\usepackage{subfigure}
\usepackage{wrapfig}
\usepackage{multirow}

\title{GPU Enhancement of the Trigger to Extend Physics Reach at the LHC}

\author{V. Halyo\thanks{Corresponding Author}, A. Hunt, P. Jindal, P. LeGresley, P. Lujan \\
\llap Princeton University, Princeton, NJ, USA \\
E-mail: \email{vhalyo@gmail.com}}

\abstract{Significant new challenges are continuously confronting the High Energy Physics (HEP) experiments, in
particular the two detectors at the Large Hadron Collider (LHC) at CERN, where nominal conditions deliver proton-proton
collisions to the detectors at a rate of 40 MHz. This rate must be significantly reduced to comply with both the
performance limitations of the mass storage hardware and the capabilities of the computing resources to process the
collected data in a timely fashion for physics analysis.  At the same time, the physics signals of interest must be
retained with high efficiency.

The quest for rare new physics phenomena at the LHC leads to evaluation of a Graphics Processing Unit (GPU) enhancement for
the existing High-Level Trigger (HLT), made possible by the current flexibility of the trigger system, which not only
provides faster and more efficient event selection, but also includes the possibility of new complex triggers that were
not previously feasible. A new tracking algorithm is evaluated on a NVIDIA Tesla K20c GPU, allowing for the first time
the reconstruction of long-lived particles at the tracker system in the trigger. Preliminary time performance and
efficiency will be presented.}

\keywords{ATLAS; CMS; Level-1 trigger; HLT; Tracker system}

\begin{document}

\section{Introduction} 

The stunning performance of the LHC, the successful physics results, and the discovery of 
a new Higgs-like particle will mark in history the first LHC running period not as the beginning of the end but rather
the end of the beginning. The first LHC shutdown, scheduled for 2 years, provides a remarkable opportunity to
improve detector performance even further; in particular, to enhance and extend the physics reach by improving the set of events selected and recorded by the trigger.

The Trigger and Data Acquisition (DAQ) systems~\cite{Sphicas:2002gg}, \cite{ATLAS:2003aa}, \cite{Chatrchyan:2008aa}, \cite{Aad:2008zzm}
in a modern collider experiment, such as the ATLAS and CMS experiments at LHC, provide essential preliminary online analysis of the raw data for the purpose of
filtering potentially interesting events into the data storage system.
At the design luminosity of $10^{34}~\mathrm{cm}^{-2}\mathrm{s}^{-1}$ this is a formidable task,
which requires real-time handling of raw data with beam crossings at $40~\mathrm{MHz}$, each yielding
on average ${\sim}20$ inelastic proton-proton interactions, and producing approximately 1 MB of zero-suppressed data.
To keep the overall data rate within the current capability of the archival storage
system of ${\sim}100~\mathrm{MB}/s$, the trigger system must achieve a rejection ratio for background
events of at least ${\sim}400,000:1$.

The required level of performance is achieved in both CMS~\cite{Chatrchyan:2008aa} and ATLAS~\cite{Aad:2008zzm}
 by implementing a typical trigger system with a hierarchy of multiple levels, ranging from fast and relatively simple criteria implemented
entirely in hardware and firmware, to more sophisticated software-based analysis. 
While the discussion in this paper focuses principally on the CMS detector, the trigger principles and performance are
comparable between the two and the discussions can be generalized to both detectors.
The CMS trigger is implemented in 2 levels: a hardware Level-1 trigger reducing the event rate
to less than $100~\mathrm{kHz}$, followed by further processing in the High-Level Trigger (HLT)~\cite{Sakulin:2007rj}
designed to reduce this maximum Level-1 acceptance rate of 100 kHz to a final output rate of 100 Hz.
The primary goal of the HLT is to apply a specific set of
physics selection algorithms on the events read out and accept the events with the most interesting physics 
content. This computationally intensive processing is executed on a farm of commercial CPU processors that
constitute the CMS HLT hardware.

By its very nature of being a computing system, the HLT relies on technologies that have evolved
extremely rapidly.  For many decades, one of the important methods for improving the performance of computing devices
has been to increase the clock speed of the processor, and the typical CPU clock speed has increased by nearly a factor of 1000.
In the past decade, however, it has become apparent that it is no longer possible to rely solely on increases
in processor clock speed as a means for extracting additional computational power from existing architectures.
The underlying reasons are complex, but center around reaching what may be fundamental limitations in
semiconductor device physics.  For this reason, recent innovations have focused around {\em parallel}
processing, either through systems containing multiple processors, processors containing multiple cores, and so on.

One interesting technology which has continued to see exponential growth is graphics processing.  
A modern Graphic Processing Unit (GPU) is a massively parallel processor with thousands of execution units to handle highly parallel workloads
related to computer graphics.  By making these execution units highly programmable, manufacturers have made
the massive computational power of a modern GPU available for more general-purpose computing,
as opposed to being hard-wired for specific graphical operations.  In certain applications that can be executed in massively parallel
 fashion, this can yield several orders of magnitude better performance than a conventional CPU.

Furthermore, leading technology vendors have released supported development APIs, such as NVIDIA
CUDA ("Compute Unified Device Architecture'')~\cite{bib:CUDA}, which allow rapid development of
software in nearly standard C code. CUDA allows CPU-based applications to access the resources of a GPU 
for more general-purpose computing without the limitations of using a graphics API.
 A modern GPU can simultaneously execute many thousands of computations in parallel, and an application using 
effective collaboration between the CPU and the GPU can see dramatic acceleration in algorithm execution.

A particularly good application for this technology would be to try executing machine vision
and pattern recognition algorithms on detector data. These algorithms are often parallel
by their very nature, and the highly controlled data produced by a particle physics detector
reduces the pattern recognition task to its purest form.  From the physics perspective,
such an enhancement of the trigger capabilities would allow inclusion of new tracking triggers;
for example, the selection of events with multiple displaced
vertices at any location in the silicon tracker, which would be a ``smoking gun'' for new topological
signatures not predicted by the Standard Model. Such a new tracking algorithm would allow
augmentation of the existing trigger with new types of trigger filters to select events with topological signatures
that are currently not recorded efficiently.

One should note that this paper does not imply that executing machine vision and pattern recognition 
algorithms is necessarily more appropriate for GPUs compared to the conventional Combinatorial Track 
Finder algorithms used in high energy physics. Rather, it is a complementary way that was first studied
 by the authors due to the similarity of the problem with computer image analysis. These techniques have
 proven successful in image processing using the Hough transform on GPUs.

In the following, a description of the physics motivation behind GPU enhancement of the HLT is provided.  This
includes a brief description of the existing High Level Trigger and the tracking algorithm used, and the GPU
and CUDA architecture.  Additional capabilities would be enabled by a hybrid GPU/CPU computing farm, and the
use of GPUs for the HLT systems of both CMS and ATLAS is being studied by several
groups~\cite{Emeliyanov:2012mg}, \cite{Mattmann:2012hi}, \cite{Clark:2011zzb}. These studies have so far,
however, principally focused on improving the speed of existing tracking algorithms rather than introducing
new tracking techniques.  The authors propose a new tracking reconstruction algorithm, show preliminary
results, and discuss proposed new triggers that could trigger on events with a highly distinct signature which
would be a clear indicator of new physics at the LHC.

\section{Physics Motivation}

An enhancement of the HLT might permit processing of the new tracking reconstruction on the full event 
at the Level-1 event rate up to design luminosity. The proposed HLT tracking would have a far-reaching impact on the physics program 
set by the HLT where the ultimate goal is to select the events of interest for CMS.
The new tracking algorithm would also introduce new possible trigger paths that are currently not possible due to the extensive
 processing time that would be required using CPUs alone. The robust parallel processing of the tracking on the 
hybrid CPU/GPU system would allow reconstruction of not only charged prompt tracks from the interaction point
but also reconstruction of displaced vertices in the tracker far beyond what is possible in both CMS and ATLAS.
Therefore, the tracking algorithm using the GPU will enrich the physics program and allow for searches on models with new topological signatures that were not possible or suppressed before.

Only a few of the large selection of topological models can be described here. An example of an
interesting class of models which would benefit from such triggers are hidden-valley models~\cite{bib:hiddenvalley},~\cite{Han:2007ae},
in which a new confining gauge group is added to the standard model. The resulting (electrically-neutral) bound states can have
 low masses and long lifetimes and could be observed at the LHC. The production multiplicities are often large and events with final states with heavy flavor are
common. In addition, displaced vertices and missing energy are possible. Accounting for LEP constraints, LHC production 
cross-sections were estimated to typically be in the 1-100 fb range, though they can be larger.

New tracking triggers would permit selection at the HLT level of events where a Higgs-like particle decays with a substantial branching fraction to long-lived neutral particles decaying
at macroscopic distances from the primary vertex~\cite{bib:hiddenvalley} in the tracker. The limited experimental constraints on light neutral long-lived particles
suggests that one should probably not be surprised if a Higgs-like particle reveals itself  through such decays. As the lifetimes of these resonances 
are not constrained, decays at centimeter and meter scales are equally possible. Each Higgs-like decay may produce two or more resonances 
with the multiplicity possibly varying from event to event. In many models the resonances will decay to the heaviest fermion 
pair available, with branching fractions similar to those of the standard model Higgs.

Other models that predict these unusual signatures~\cite{bib:hiddenvalley}--\cite{Strassler:2006ri} include either simple models where a scalar is added to the Higgs potential or superpotential depending on whether a nonsupersymmetric
or supersymmetric framework was used to build the model. Another possibility is confining hidden valley models, which produce qualitatively similar signals, though the origin of the signals is 
quite different.  A complex Higgs decay in a hidden-valley model could produce, for example, four resonances: one decaying promptly to jets,
 one escaping the detector giving missing energy, and two decaying to $b\bar{b}$, each with a displaced vertex.  

Other obvious channels that would benefit from the new tracking algorithm allowing triggers on displaced jets
or vertices are inclusive and exclusive $b$-decay channels, or various topologies with boosted jets that will
be more frequent as the center of mass energy of the collision at design luminosity is increased. These new
triggers will naturally be sensitive to a larger parameter space, including lower-mass Higgs, that could have
evaded detection in previous experiments. For example, CMS has performed a search for long-lived neutral
particles decaying to displaced leptons~\cite{Chatrchyan:2012jna}, but the efficiency for measuring leptons
originating from a low-mass Higgs of $M_H = 120$ GeV/$c^2$ is very low due to the lepton momentum requirements
in the trigger, resulting in relatively poor limits for this mass region.

\section{Trigger System}

The LHC delivers proton-proton collisions to the CMS~\cite{CMS:1994hea} detector at a rate of 40~MHz. This rate
must be significantly reduced to comply with the performance limitations of the mass-storage hardware and the ability
of the offline computing resources to process the collected data in a timely fashion for physics analysis. At the same
time, the physics signals of interest must be retained with high efficiency. CMS features a two-level trigger system to reduce
the rate to approximately 100~Hz. The Level-1 trigger~\cite{Dasu:2000ge} is based on custom hardware and designed to reduce the rate
to about 100~kHz, corresponding to 100~GB/s, assuming an average event size of 1~MB. The High Level Trigger (HLT)~\cite{Sphicas:2002gg},\cite{Sakulin:2007rj}
is purely software-based and must achieve the remaining rate reduction by executing sophisticated offline-quality algorithms.
An HLT path consists of algorithms that are executed in order of increasing complexity, so that
the execution of a path can be stopped quickly as soon as evidence for the signal of interest is found to be missing.
This optimization means that the more sophisticated and time-consuming reconstruction algorithms are only applied when necessary.
The ability to select the events of interest is the foundation of a
quest for rare new physics phenomena, and thanks to the flexibility of the CMS HLT software and hardware computing farm an enhancement for the HLT is proposed.

\begin{figure}[!Hhtb]
  \begin{center}
    \includegraphics[width=1.0\textwidth]{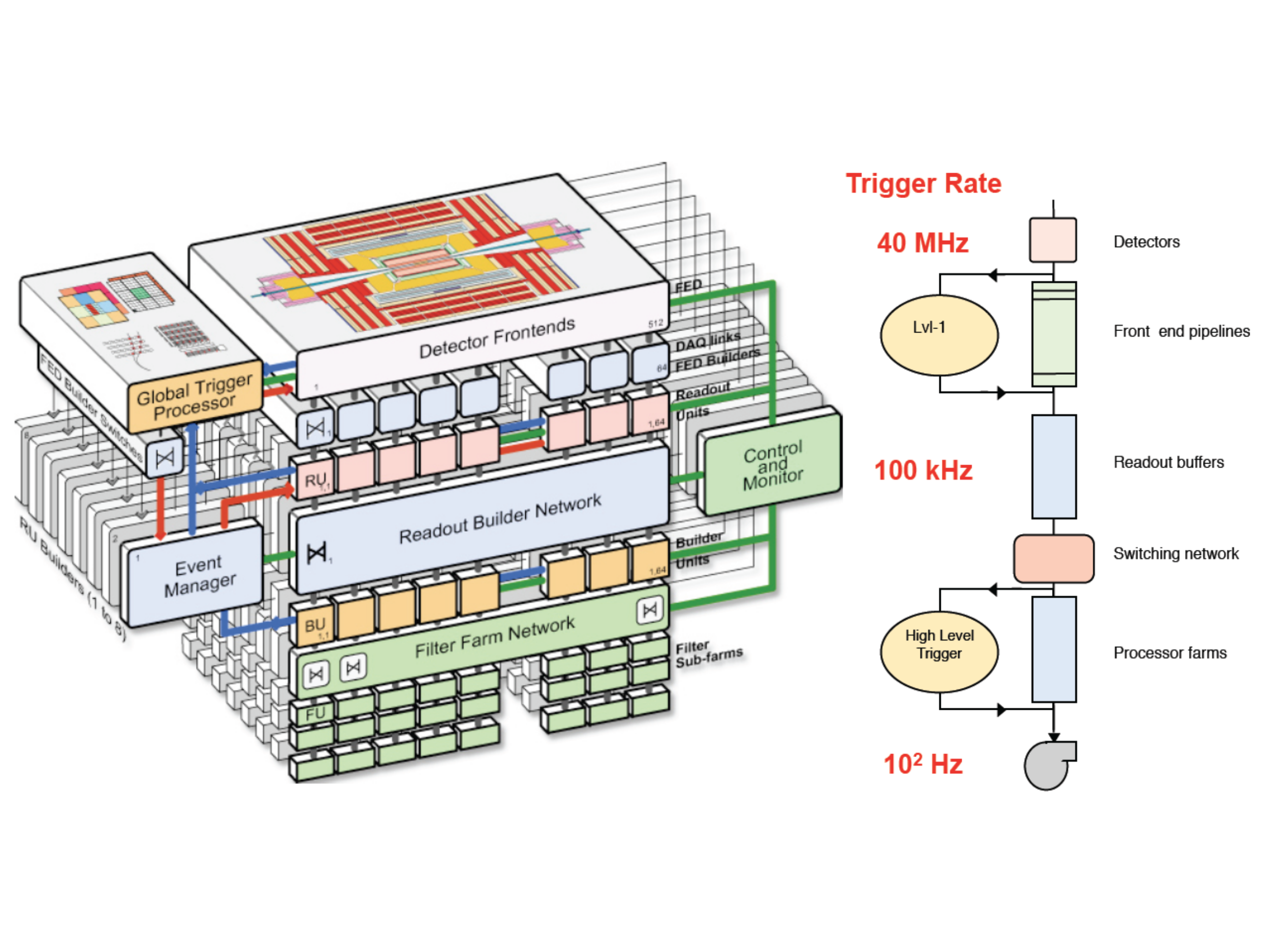} \label{fig:hltarc}
    %\subfigure[]{\includegraphics[width=0.5\textwidth]{figs/Tesla_M1060_1u_Super_Micro_Front_Elevated_no_cover_large.pdf} \label{fig:integratedsys}}
      \caption{CMS DAQ architecture. The size of the event builder (72 Readout Units, 288 Builder Units) represents one slice; the system can be equipped with up to eight slices.}
      %Bottom: Integrated GPU/CPU 1U server (image courtesy of NVIDIA).}
  \end{center}
\end{figure}

% for VH -- reference bib:datamodel is missing here
The architecture of the CMS HLT~\cite{Sphicas:2002gg},\cite{Sakulin:2007rj} combines the flexibility gained by using the offline
reconstruction~\cite{bib:datamodel} with the robustness required for reliable online operation of the DAQ. Figure~\ref{fig:hltarc} shows 
a schematic of the architecture of the CMS DAQ system. Event fragments are read out and stored in Readout Units (RU) for each event accepted
by the Level-1 trigger. The fragments are subsequently assembled into complete events by an Event Builder through a complex of switched 
networks into event buffers (BU). The full event content is then handed to one of the HLT Filter Units (FU). The FU execute a series of 
physics reconstruction and filter algorithms and events that are found to be sufficiently interesting for offline analysis are forwarded 
to the Storage Manager (SM). The decoupling of physics algorithm execution from data flow allows each FU to continue operation in the case of a problematic event, recover the
content of the problematic event, and forward it to be stored unprocessed. The FU architecture consists of 
two separate applications, the ResourceBroker, which exchanges data with the DAQ, and the EventProcessor, which integrates the reconstruction
software. Given the involved event sizes and rates, reformatting raw events consisting of a large number of small data fragments into the physical 
memory of a Filter Unit requires high bandwidth I/O, while the subsequent HLT processing is mostly CPU-intensive.
The EventProcessor previously mentioned encapsulates the event processing machinery of the CMS reconstruction, providing the full flexibility 
required to execute complex physics algorithms. 

As the farms of machines where these event selections are executed are commercial servers, GPUs can be easily integrated into them.
A simple solution is 1U servers containing dual multicore CPUs and two NVIDIA Tesla GPUs. 
While this would provide a single, integrated 1U GPU/CPU server, this of course implies replacing the existing servers, 
evaluating the new power consumption, and the cost of these servers relative to any other options.

\section{GPU Architecture}

The NVIDIA GPU architecture consists of a
number of general purpose floating point processors: {\it stream processors}.
The stream processors are fully programmable and can be utilized as general
purpose computing cores for computationally intensive purposes outside of graphics.
Multiple stream processors, a parallel data cache (a user managed
on-chip cache), and, on some processors, a hardware-managed L1 cache are
grouped together into a multiprocessor.  Typical high end GPUs feature tens
of multiprocessors with hundreds of stream processors.

The streaming processors can be utilized for general purpose computations using
CUDA ("Compute Unified Device Architecture")~\cite{bib:CUDA}.  Unlike past efforts to harness
the computational power found in GPUs using a graphics API, CUDA specifies a
few simple extensions to the C programming language and a parallel programming
model to enable full access to the GPU hardware.

The programmer writes a CPU program that invokes functions called {\it kernels}
that execute on the GPU.  A kernel consists of many thousands of threads that
execute in parallel across the stream processors.  Groups of threads called
{\it blocks} are scheduled on a single multiprocessor and can cooperate among
themselves through access to a parallel data cache.  Each block is independent
by design so the many blocks of a kernel, called a {\it grid}, can be executed
in parallel on several multiprocessors.

\section{Inner Tracker}

\begin{figure}[!Hhtb]
\begin{center}
	\includegraphics[width=1.0\textwidth]{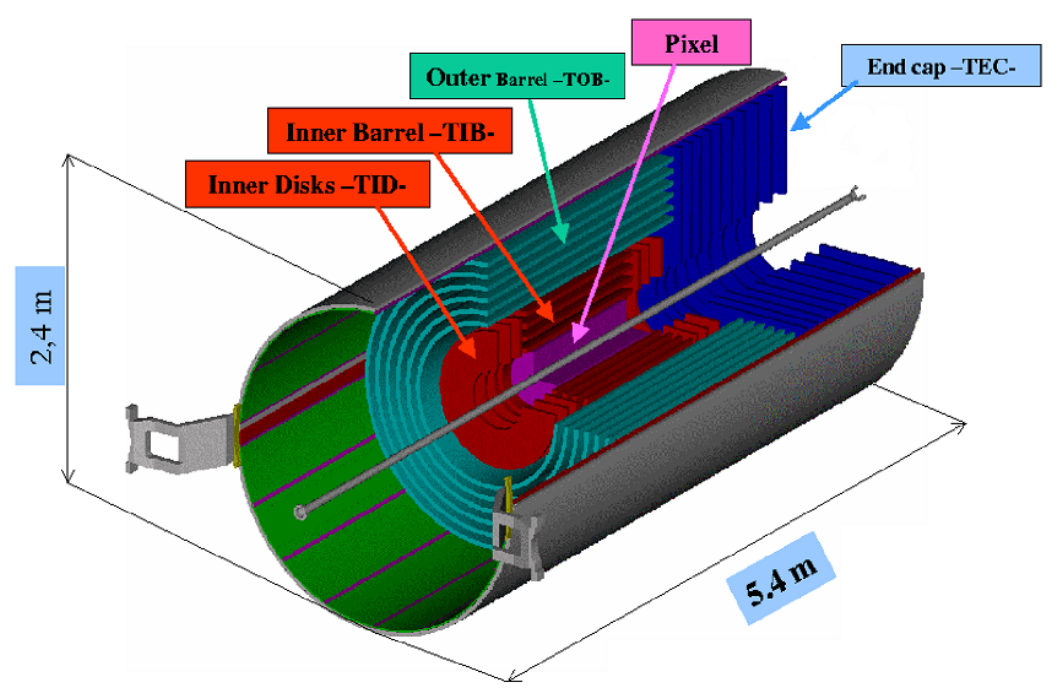}
	\includegraphics[width=1.0\textwidth]{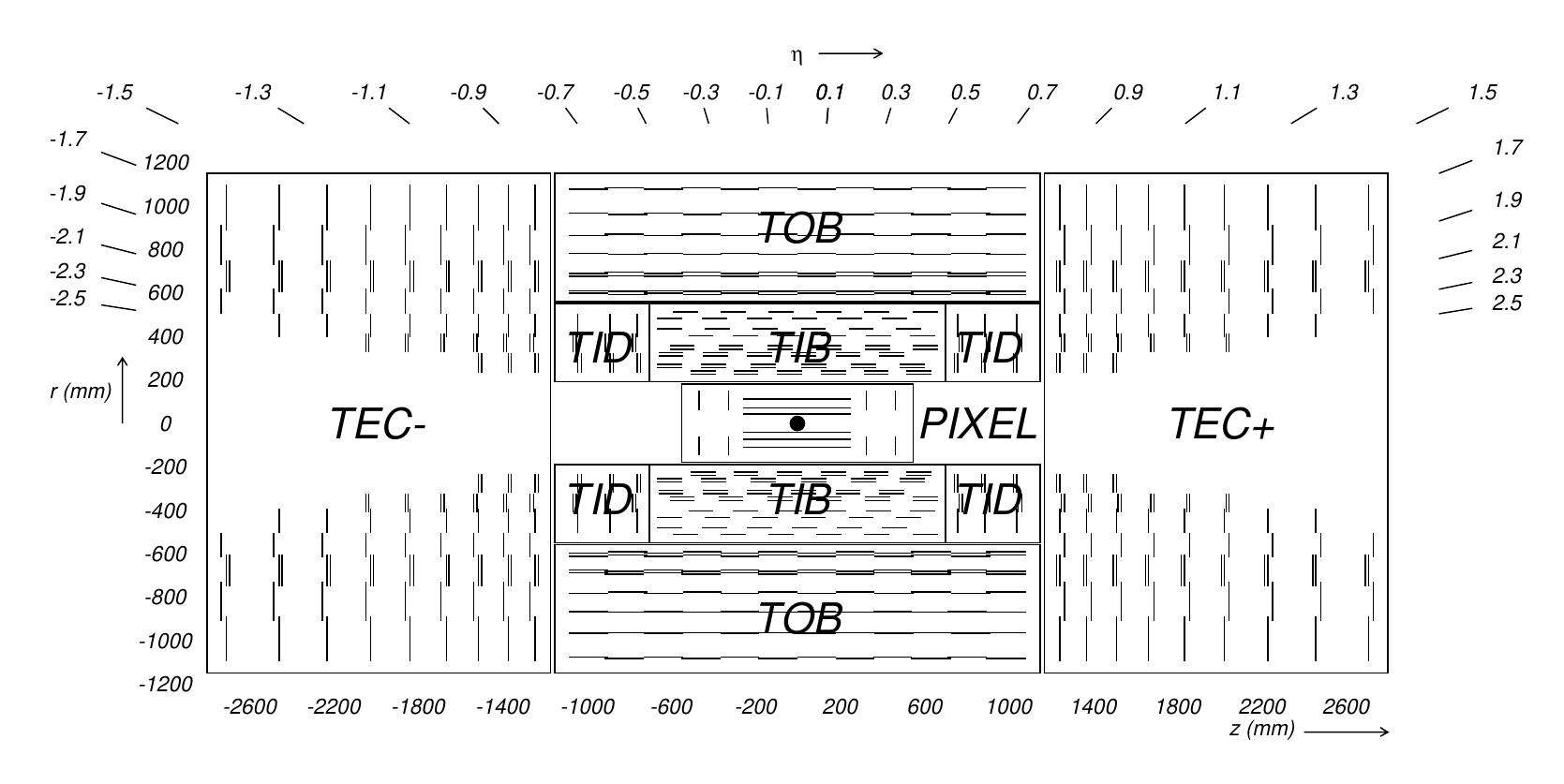}
	\caption{The CMS tracker in a 3-D view (top) and a 2-D view (bottom), showing the pixel detector and the silicon
	strip tracker. \label{fig:tracker}}
	\end{center}
\end{figure}

Both the CMS and ATLAS inner trackers include a silicon pixel detector and a silicon strip detector.
All tracker layers provide two-dimensional hit position measurements, but only the pixel tracker
 and a subset of the strip tracker layers provide three-dimensional hit position measurements.
Only the CMS tracker is considered; however, the detector performances are comparable
in the CMS and ATLAS experiments, and the tracking algorithm and results discussed are applicable to both of 
these experiments. The CMS pixel detector includes three barrel layers and two forward disks
on either end of the detector; outside the pixel detector is the strip detector, consisting of ten layers in the barrel plus three 
inner disks and nine forward disks at each end of the detector.
Owing to the strong magnetic field and the high granularity of the silicon tracker, 
promptly-produced charged particles with transverse momentum $p_T = 100$ GeV/c are reconstructed 
with a resolution in $p_T$ of 1.5\% and in transverse impact
parameter $d_0$ of 15 mm. The track reconstruction algorithms are able to reconstruct displaced
tracks with transverse impact parameters up to 25 cm from particles decaying up to 50 cm
from the beam line. The performance of the track reconstruction algorithms has been studied
with data~\cite{Khachatryan:2010pw}. The silicon tracker is also used to reconstruct the primary vertex position with
 a precision of $\sigma_d = 20$ mm in each dimension. Figure~\ref{fig:tracker} shows the layout of the CMS tracker.

\section{Current Tracking Algorithm}

The CMS track reconstruction, known as the Combinatorial Track Finder (CTF), uses an iterative algorithm,
with earlier iterations searching for tracks that are more easily found (relatively higher $p_T$ tracks
close to the interaction region). The hits in these tracks are then removed, allowing later iterations to
search for lower momentum or highly displaced tracks without the combinatorial possibilities becoming too
large. Each iteration consists of four steps: a seeding step, a track finding step, a track fitting step,
and a selection step.

In the first step, a seed, consisting of two or three hits and a primary vertex constraint,
is constructed to create an initial estimate of the trajectory parameters for the track. Second, the
track finding is performed using a global Kalman filter~\cite{Fruhwirth:1987fm}. This step performs a fast
propagation of the track candidate to propagate the track through the layers of the detector, search for
compatible nearby hits, and attach them to the candidate. After all candidates in this step have been
found, the track candidate collection is then cleaned to remove duplicate tracks or tracks which share a
large number of hits.  The third step is to perform a full Kalman fit over the whole track to obtain the
best estimate of the track parameters at all points along the trajectory. The filter begins at the
innermost hits, and then iterates outward through each hit to update the track trajectory estimate and
its uncertainty. After this first fit is complete, a smoothing stage is then performed running backwards
from the last hit to apply the information from the later hits to the earlier ones. This step uses a
Runge-Kutta propagator to account for the effect of material interactions and an inhomogeneous magnetic
field. Finally, track selection requirements are applied to reduce the fake rate of the resulting track
candidates.

\section{Proposed Fast Tracking Algorithm}

One possible application of the GPU-enabled parallelism would be to accelerate the performance of the
existing combinatorial track finder.  Since the finding and fitting process for each track is largely
independent of the other tracks, parallelization should not be conceptually difficult.  However, the use
of GPU-based computing in the HLT also enables the possibility of running tracking algorithms which are
dramatically and qualitatively different in nature to the current CTF, thus enhancing the discovery potential.

\begin{figure}[!Hhtb]
\begin{center}
  \subfigure[]{\includegraphics[width=0.32\textwidth]{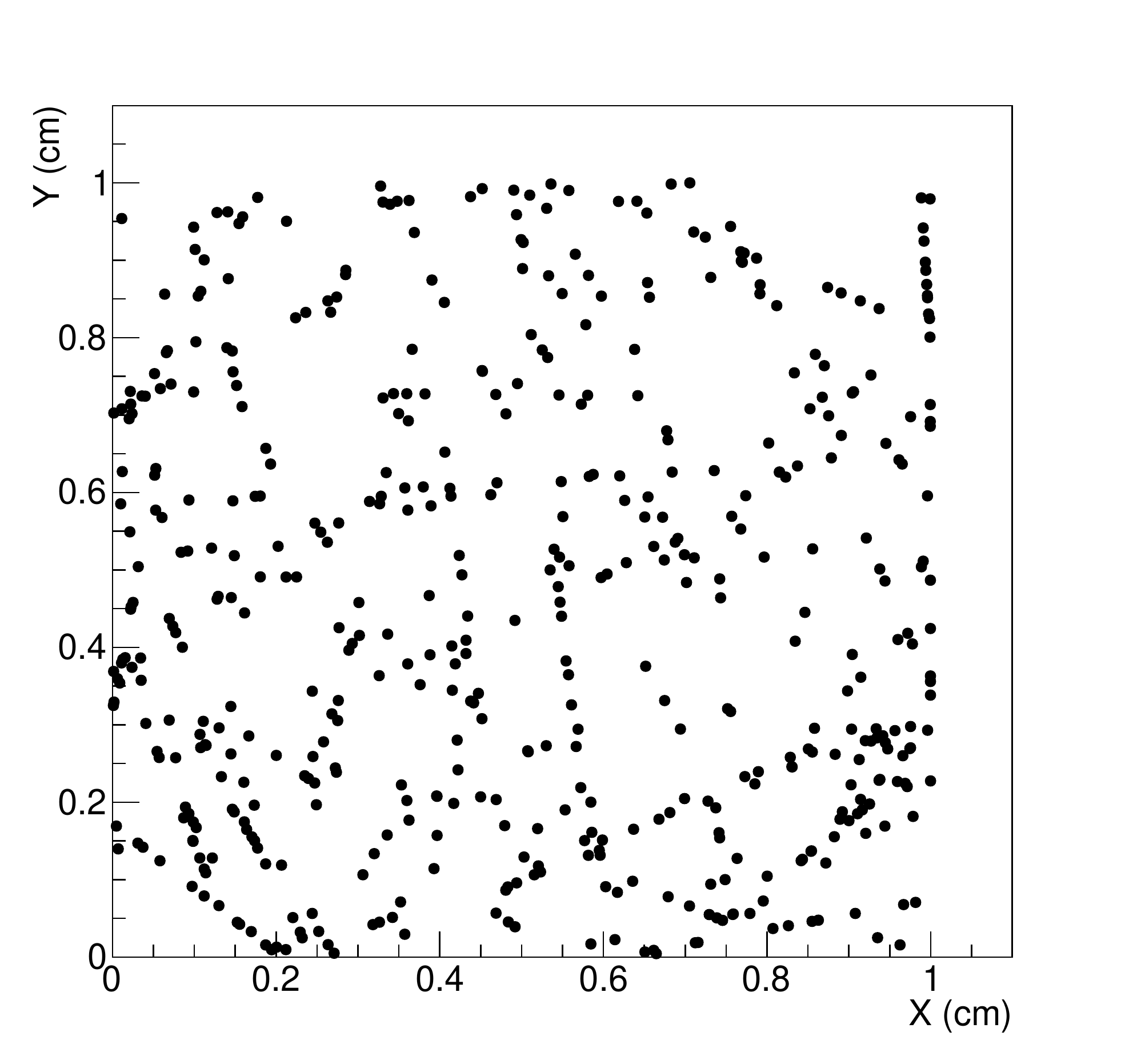} \label{fig:hits}}
  \subfigure[]{\includegraphics[width=0.30\textwidth]{figs/50events_accumulator.pdf} \label{fig:accumulator}} 
  \subfigure[]{\includegraphics[width=0.32\textwidth]{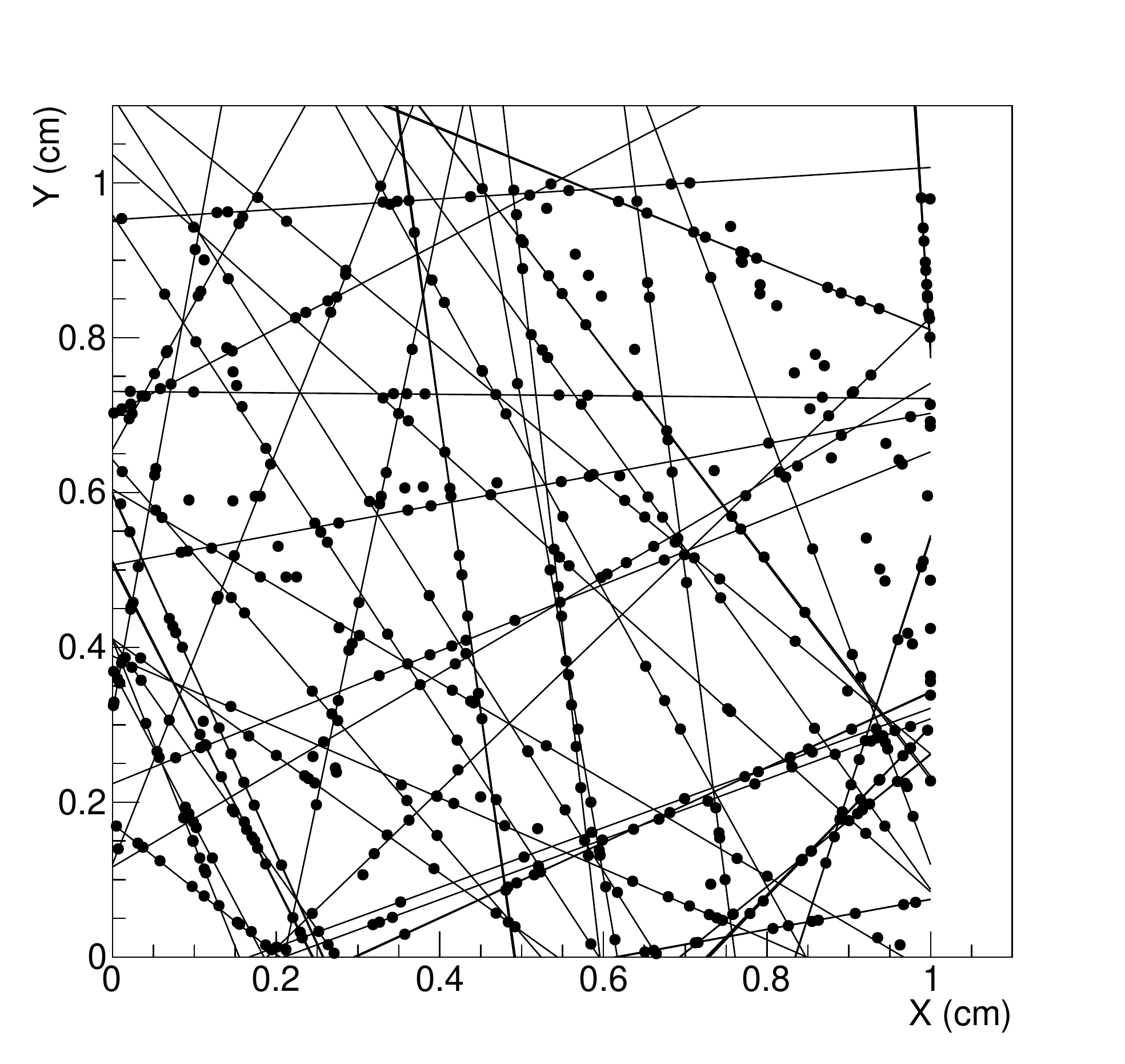} \label{fig:tracks}}
  \caption{The Hough transform algorithm applied to a simple example. Left: Hits in a simulated event with 50
    straight-line tracks and 10 hits per track. Center: Each hit results in a sinusoidal curve in the parameter
    space. Locations where many of these sinusoidal lines cross are likely to be tracks in the original data. Right:
    Candidate tracks identified from finding local maxima in the parameter space.\label{fig:hough}}
\end{center}
\end{figure}

 The Hough transform is a well-known algorithm used in various machine
vision applications that differs from the CTF approach in that it does not operate on
localized features of a data set.  Rather, the technique is in some sense more
holistic, operating on an entire image as a whole. 
The technique is described at length in existing literature~\cite{bib:HT1},~\cite{bib:HT2},~\cite{bib:HT3},
but the concept as it would be applicable to the
tracking problem is to consider the parameterization of any given track.  Any given set of hits
observed in the detector, for example, the simulated straight line track data shown in Figure~\ref{fig:hits},
can correspond to many different possible tracks in parameter space as shown in Figure~\ref{fig:accumulator}.
When integrated over the entire data set, peaks appear in parameter space at the values corresponding to the actual, 
physical trajectories as shown in Figure~\ref{fig:tracks}.

One should note, the Hough transform approach as presented above is computationally expensive.
Hence, variations of the standard Hough transform method ~\cite{bib:LHT},~\cite{bib:AHT} 
were previously used for the purpose of track finding in high energy physics~\cite{bib:slac}-~\cite{bib:cosmicmu}.

\section{Preliminary Results}

As was briefly discussed previously, the traditional combinatoric track finder approach is constrained by the need to
keep the number of possible track combinations under control, which means that displaced vertices and other odd tracks
are often not reconstructed due to the extensive processing time required to consider them.
It is hoped that a more holistic algorithm can be used to supplement the existing
tracking algorithm in a significant way.  As a demonstration of this, a simple tracking algorithm based on
the Hough transform was developed and run against a standalone Monte Carlo simulation.
Simulations are performed using a simple detector model in which only the transverse plane is considered. The detector model 
contains a simulated beampipe with a radius of 3.0 cm, surrounded by ten concentric, evenly-spaced tracking layers with the last layer at a 
radius of 110.0 cm. Resolution for a single hit is taken to be 0.4 mm in each direction.
Performance was evaluated using Intel multicore CPUs and NVIDIA Tesla C2075 and K20c GPUs.

The Hough transform algorithm is not trivial to implement in parallel because the computation of the parameter space is
subject to race conditions, where it is not necessarily safe to have multiple threads arbitrarily make updates as hits
are processed.  The Intel Performance Primitives (IPP) library contains an implementation of the Hough transform, but as
it is not parallelized, it would be difficult to compare this fairly to a GPU-based implementation. In order to provide
a more fair performance comparison, a parallel implementation of the Hough transform for the CPU has been developed as
part of this work.  This CPU implementation is parallelized using OpenMP threads and speedups of 2.6 -- 3.6 were
achieved on a Intel Core i7-3770 (Ivy Bridge, quad core, 3.4 GHz) processor.  The CPU used for the performance results
is physically a quad core CPU but supports up to two threads per physical core using Simultaneous Multithreading (SMT),
with the Intel implementation of SMT known as HyperThreading (HT).  SMT did not show any performance advantage for this
application, because the speedup actually dropped as the number of threads exceeded the number of physical
cores.

\begin{figure}[!Hhtb]
\begin{center}
  \subfigure[]{\includegraphics[width=0.45\textwidth]{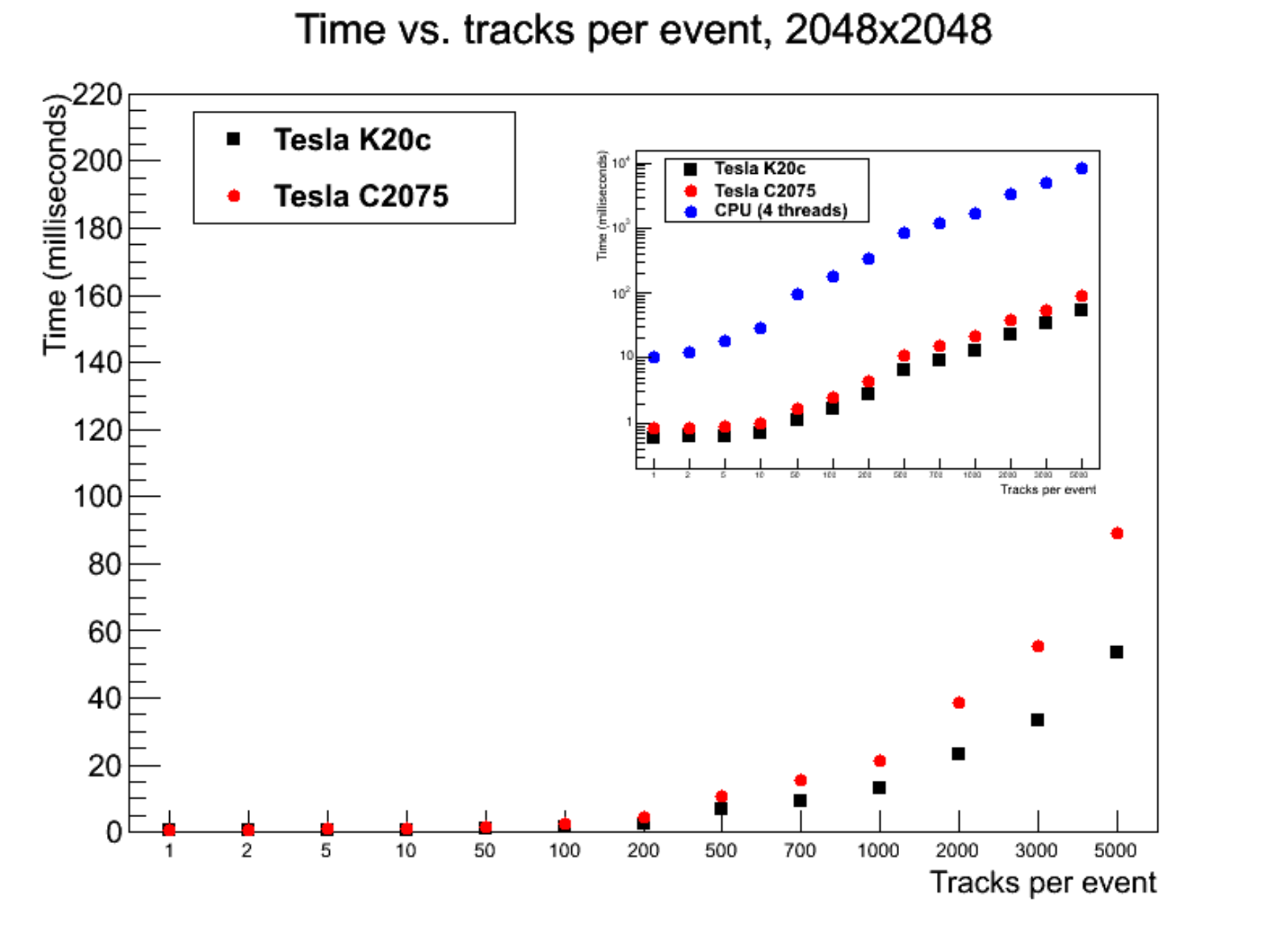} \label{fig:TimePerformance}}
  \subfigure[]{\includegraphics[width=0.45\textwidth]{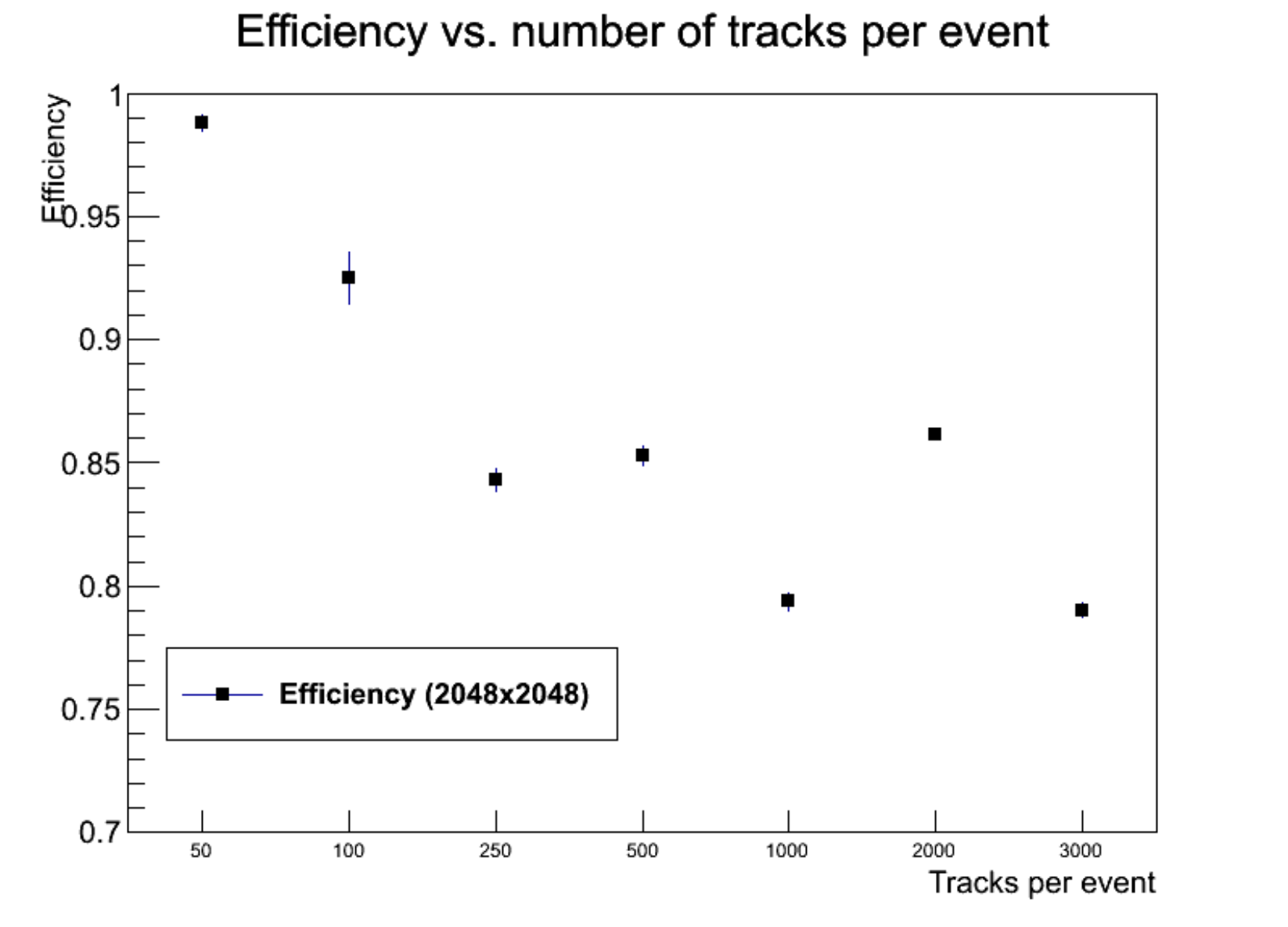} \label{fig:Eff}}
  \caption{Left: Time performance comparison of the Intel CPU (using 4 threads), NVIDIA Tesla C2075, and NVIDIA
    Tesla K20c. The inset shows the same data on a logarithmic scale. Right: Efficiency of the track-finding
    algorithm on the GPU implementation as a function of the number of tracks present in the
    event.}
\end{center}
\end{figure}

Performance results for the CPU compared to two NVIDIA GPUs are shown in Figure~\ref{fig:TimePerformance}.
The GPUs have multiple GB of memory and all of the data associated with a single event (up to 5000 hits, the
parameter space, and peak detection) reside within that memory.  Because all of this data fits in the GPU
memory with ample margins, it is not necessary to perform expensive movements of the data back and forth
between the CPU and GPU over PCI Express during the processing of an event.  For implementation in a
production environment each GPU would process a single event, and then be assigned the processing of another
event, and so on.  This means that through a combination of software and hardware support the GPU can
simultaneously be processing one event, transferring data from the CPU memory to the GPU memory for the next
event to be processed, and returning the results from the GPU memory to the CPU memory for the previously
processed event.  Since this completely overlaps data transfer with computation the timing results in
Figure~\ref{fig:TimePerformance} are for the processing of one event and exclude data transfer that would be
overlapped with the processing of other events.  In other words these timings represent the steady state
throughput of a production system and not the latency for processing a single event.
Evaluation of the memory requirement for the GPU in a case with an extreme pileup scenario of 100 
interactions per bunch crossing shows that, if one expects about 5000 tracks with an average of 15 
hits per track, this will require approximately 4.5 MB of memory using double precision. This amount 
of memory is far less than the 6 GB available in the current GPU, so we conclude that the GPU could 
easily be incorporated in the LHC DAQ.

Figure~\ref{fig:Eff} shows the efficiency defined as the fraction of simulated tracks successfully
reconstructed by the algorithm as a function of the number of tracks present in the event. As this is only a preliminary 
implementation, the efficiency and purity are not yet optimized, with the efficiency about 80\% for events with a few 
thousand tracks and comparable values for the purity. More work is obviously needed to improve the performance of the 
algorithm.

\begin{figure}[!Hhtb]
\begin{center}
	\includegraphics[width=0.45\textwidth]{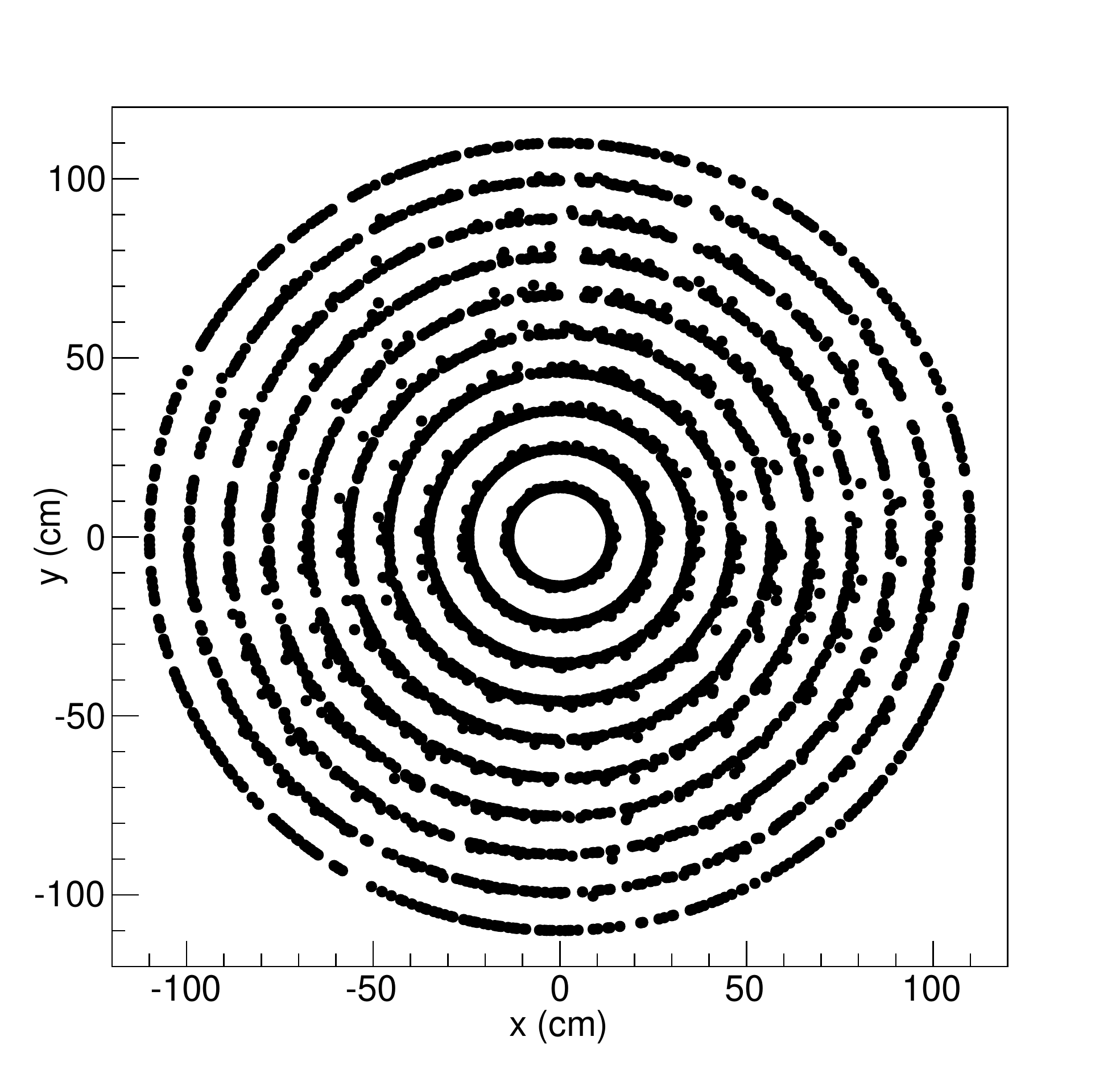}
	\includegraphics[width=0.45\textwidth]{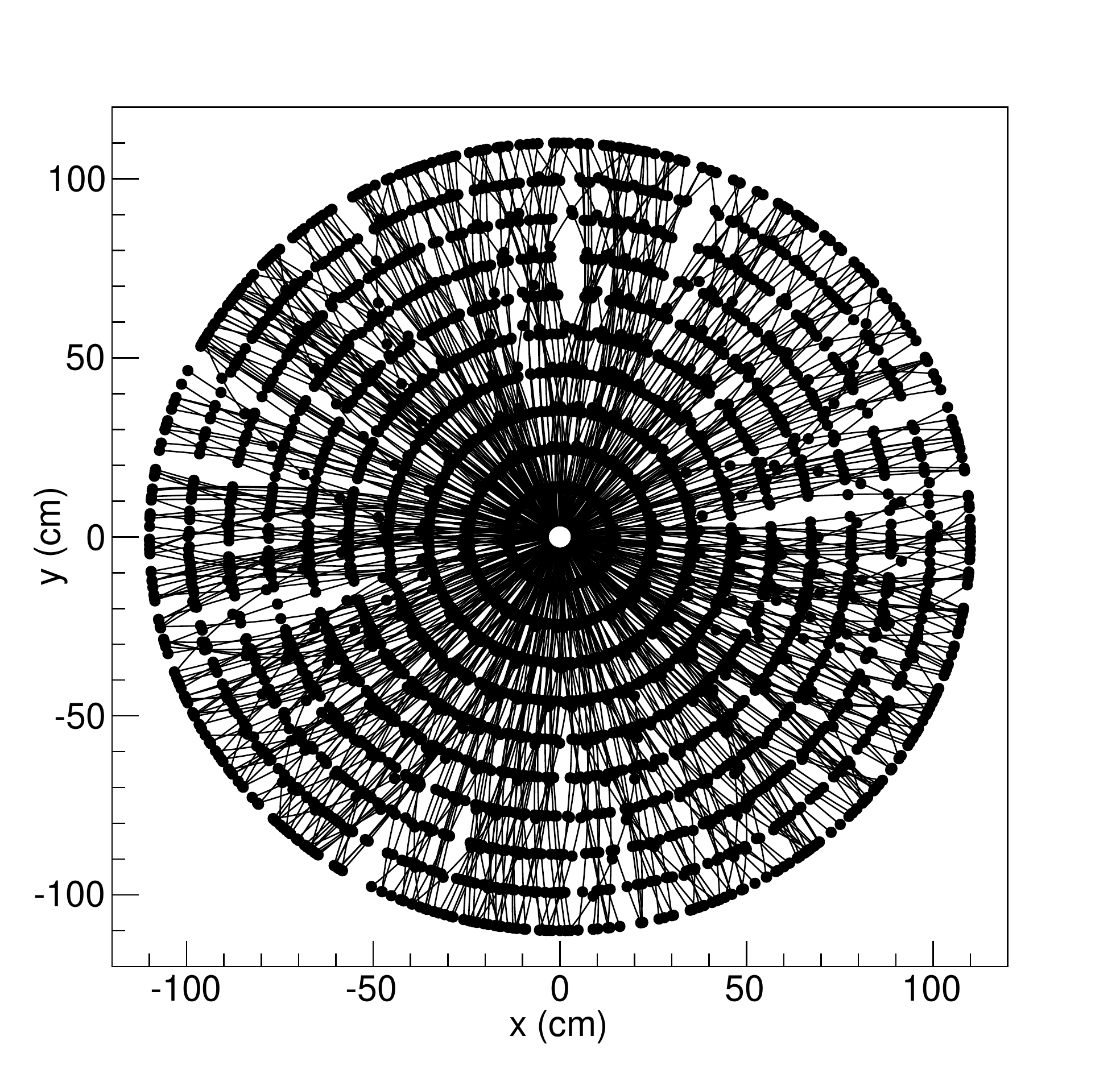}
	\caption{The Hough transform algorithm applied to a case with curved tracks. Left: Hits from Monte Carlo
	simulation for curved tracks originating from the interaction point. Right: Curved tracks identified using the
	Hough transform (86\% efficiency). \label{fig:hough_curved}}
\end{center}
\end{figure}

Although the Hough transform is most often associated with finding straight lines in data, it is applicable to any feature which can be represented 
by a finite number of parameters.  For example, circles of unknown location and radius could be identified in a three-dimensional parameter 
space representing the x and y location of the center plus a third parameter for the radius.  While the details of the particular parameterization
 have essentially no effect on the performance, the computational cost of the Hough transform is highly dependent on the total number of parameters.
Therefore, where possible the number of parameters should be minimized to reduce the computation time.  Figure~\ref{fig:hough_curved} shows
an example of results for the Hough transform implementation used for identifying curved tracks.  In this case, the
curved tracks are constrained to pass through the interaction point, reducing the parameter space to a two-dimensional space
similar to that used for straight tracks. This algorithm produces an efficiency about 86\% and comparable purity.  Curved, displaced tracks would require a
 three-parameter space, which has not yet been implemented and optimized.

\begin{figure}[!Hhtb]
\begin{center}
	\includegraphics[width=0.45\textwidth]{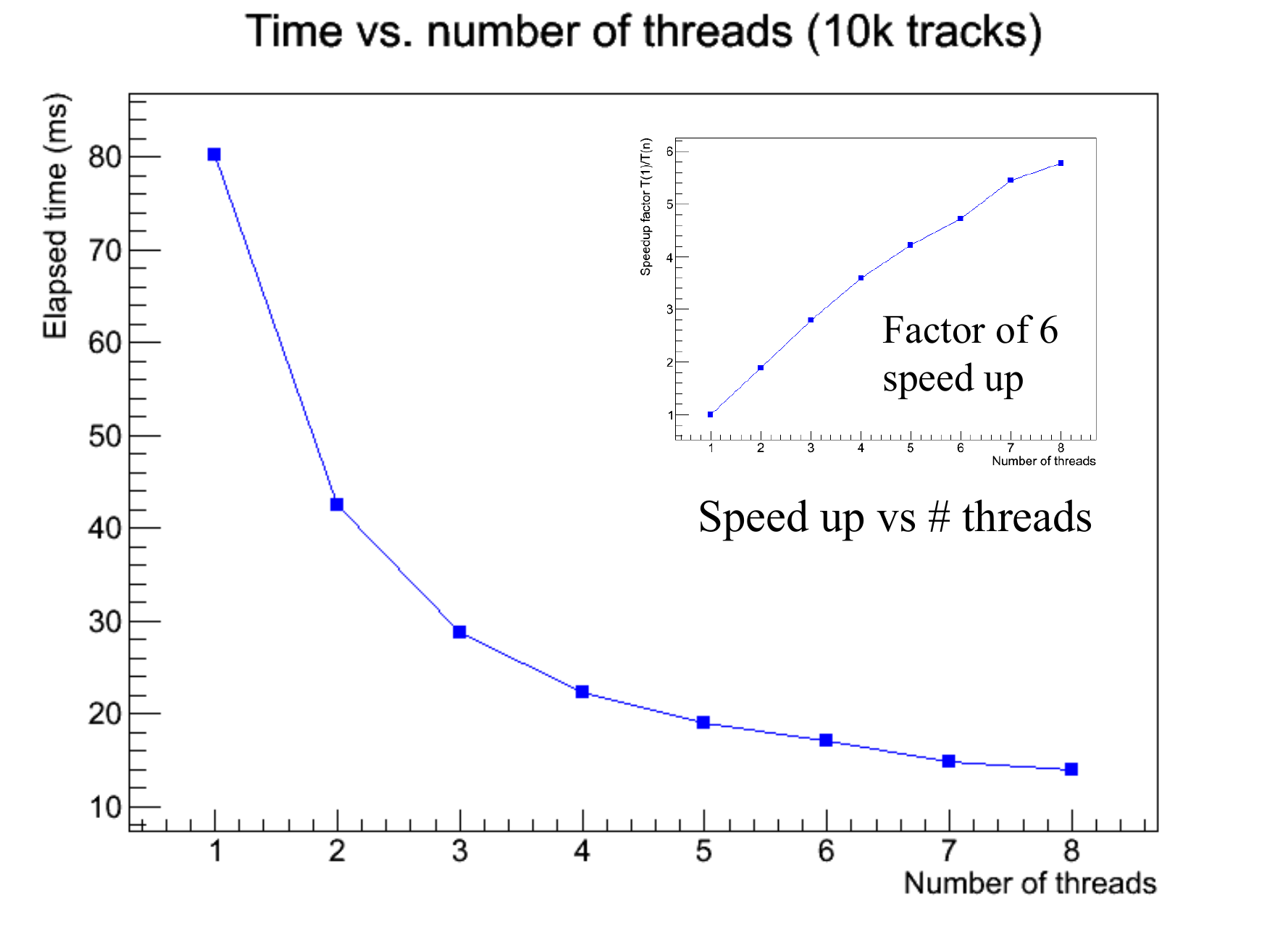}
	\includegraphics[width=0.45\textwidth]{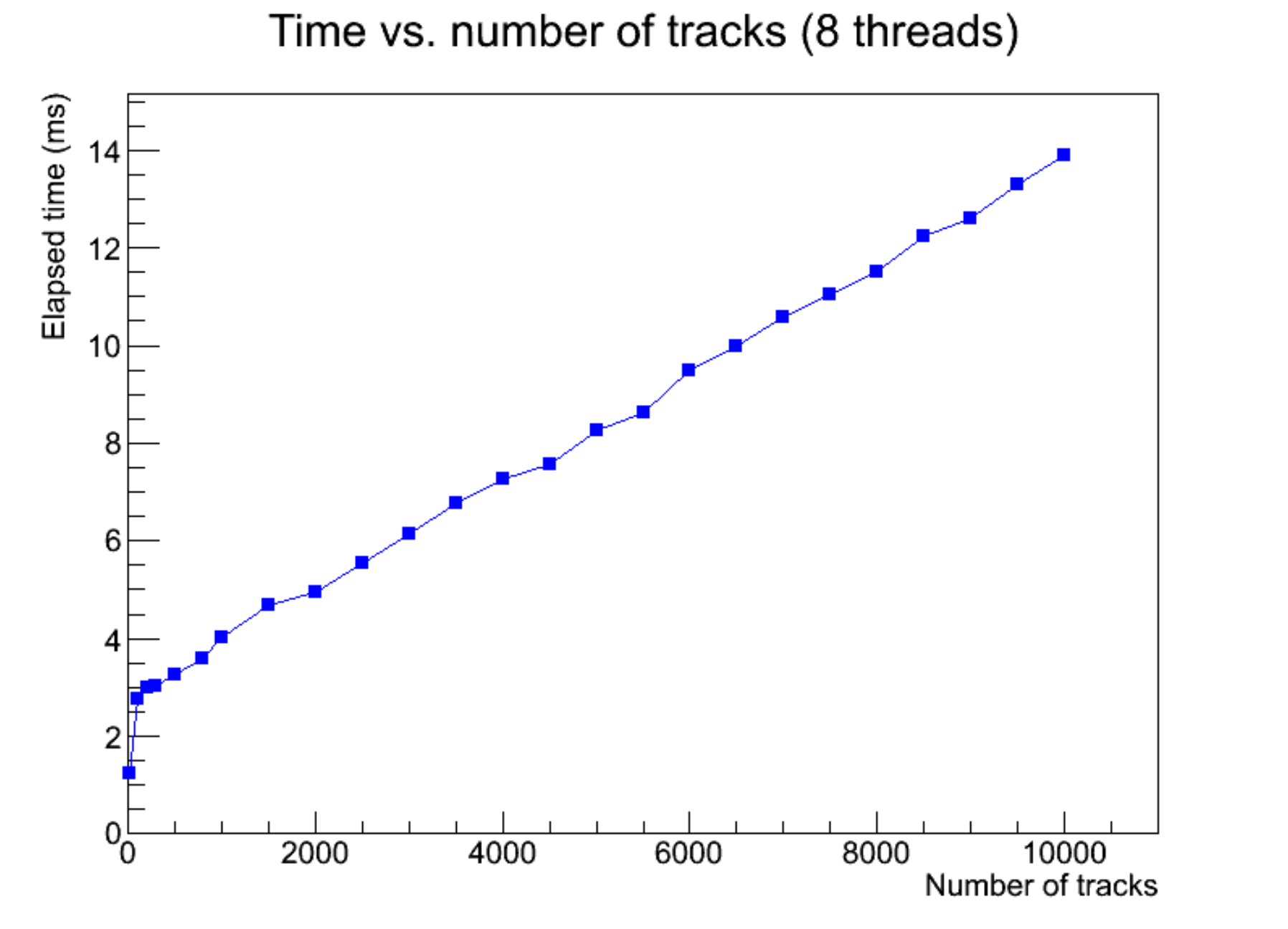}
	\caption{Time performance of a CPU-based algorithm for the fitting step of the tracking. Left: Speed-up observed
	as a function of number of threads used for the fitting. Right: Time required as a function of the number of
	tracks present in the event. \label{fig:FittingTime}}
\end{center}
\end{figure}

While the Hough transform is a critical part of the proposed tracking algorithm, it is not the only aspect of the computation.  After candidate tracks have been 
identified from analysis of the parameter space, it may be necessary to perform a fit of the hits to the candidate tracks.  This fitting operation is well suited
 to parallelization on the GPU, but the performance results in Figure~\ref{fig:FittingTime} show that a multithreaded CPU implementation
 requires only a few milliseconds.  In comparison the Hough transform on the GPU requires many tens of milliseconds, so the fitting operation is relatively low priority for performance optimization.  If or when fitting becomes a significant performance bottleneck, it could certainly be run on the GPU to further increase performance.

\section{New Trigger Capabilities}

\begin{figure}[!Hhtb]
\begin{center}
  \subfigure[]{\includegraphics[width=0.35\textwidth]{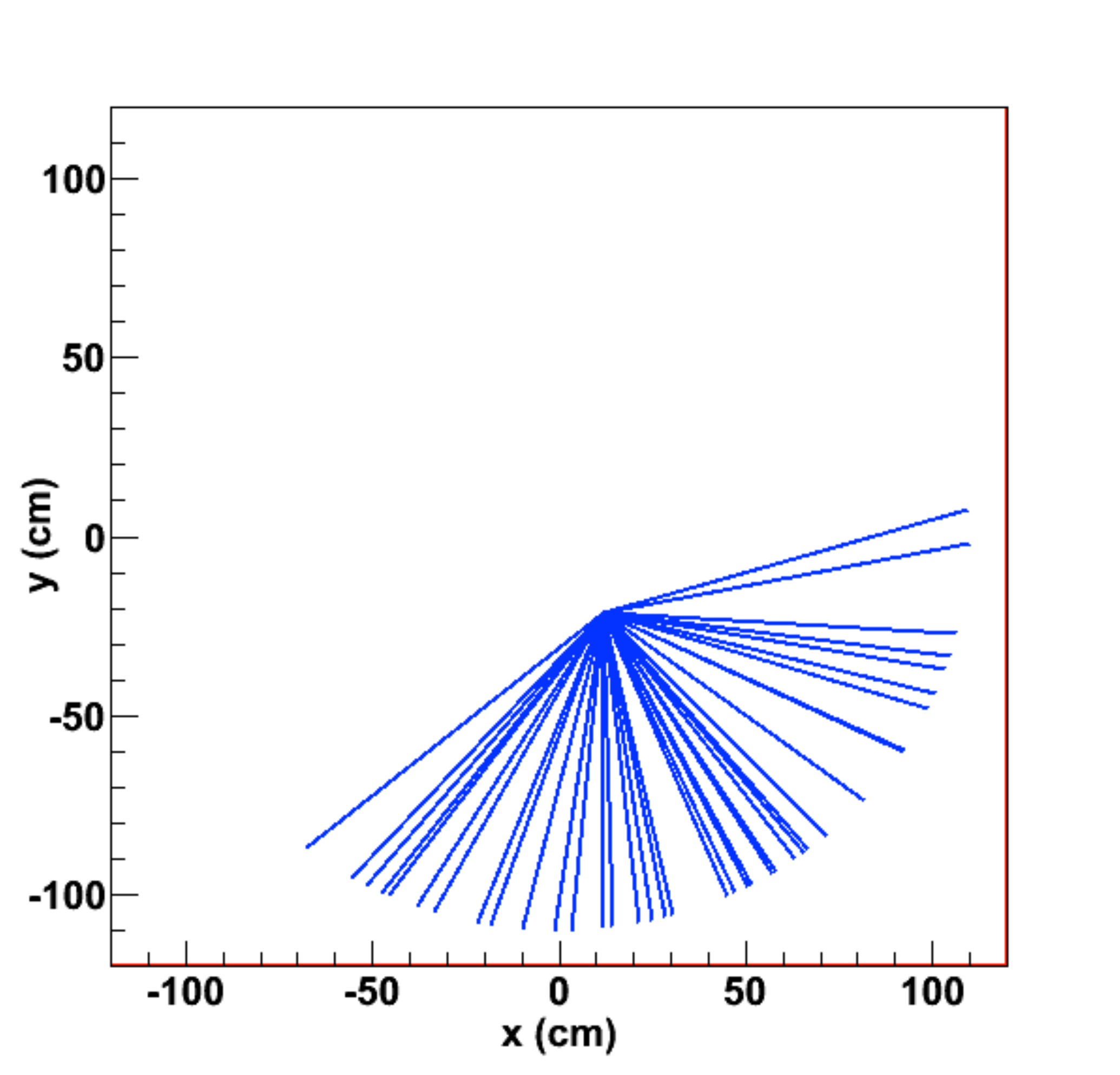} \label{fig:DisplacedJetTracks}}
  \subfigure[]{\includegraphics[width=0.30\textwidth]{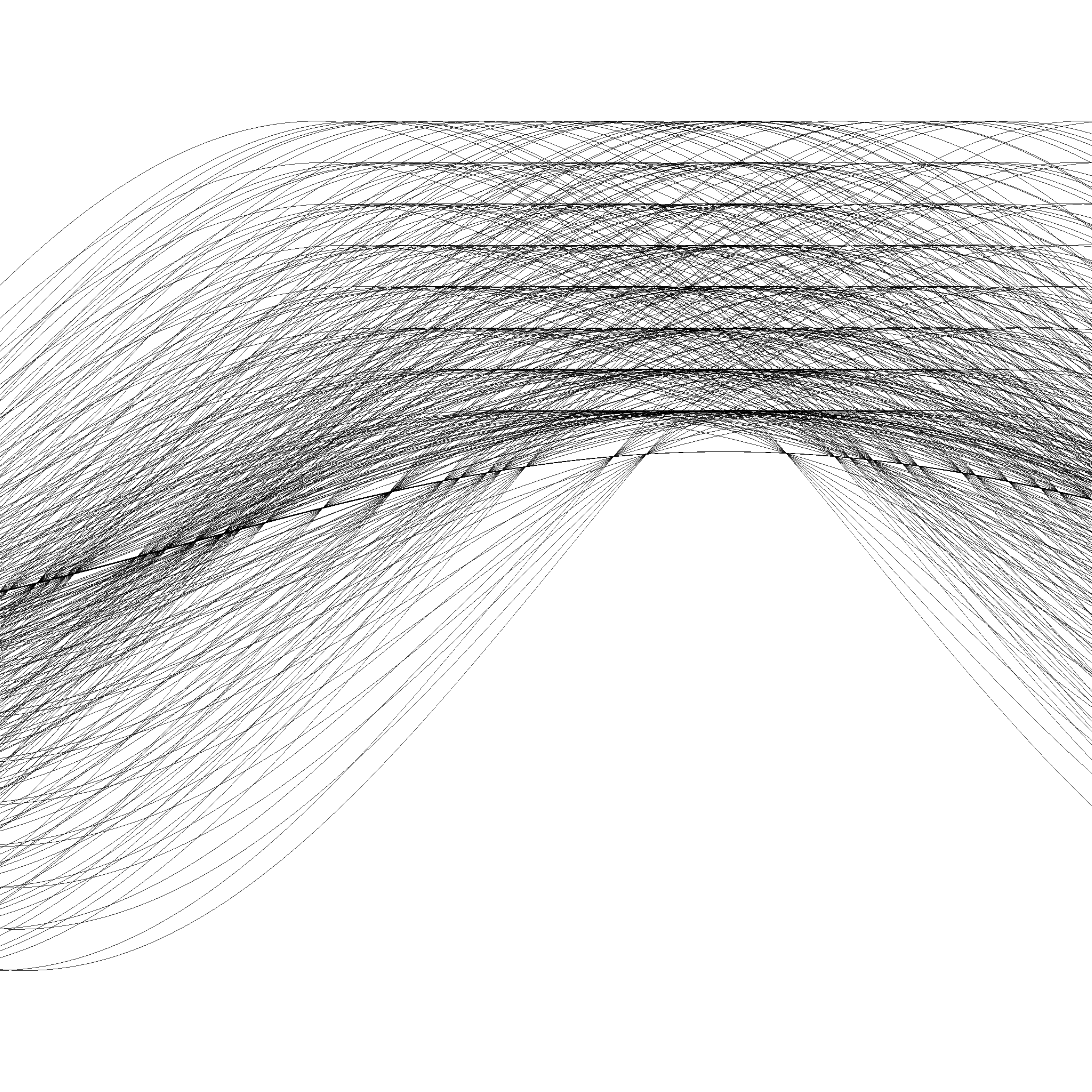} \label{fig:DisplacedJetAccumulator}}
  \subfigure[]{\includegraphics[width=0.30\textwidth]{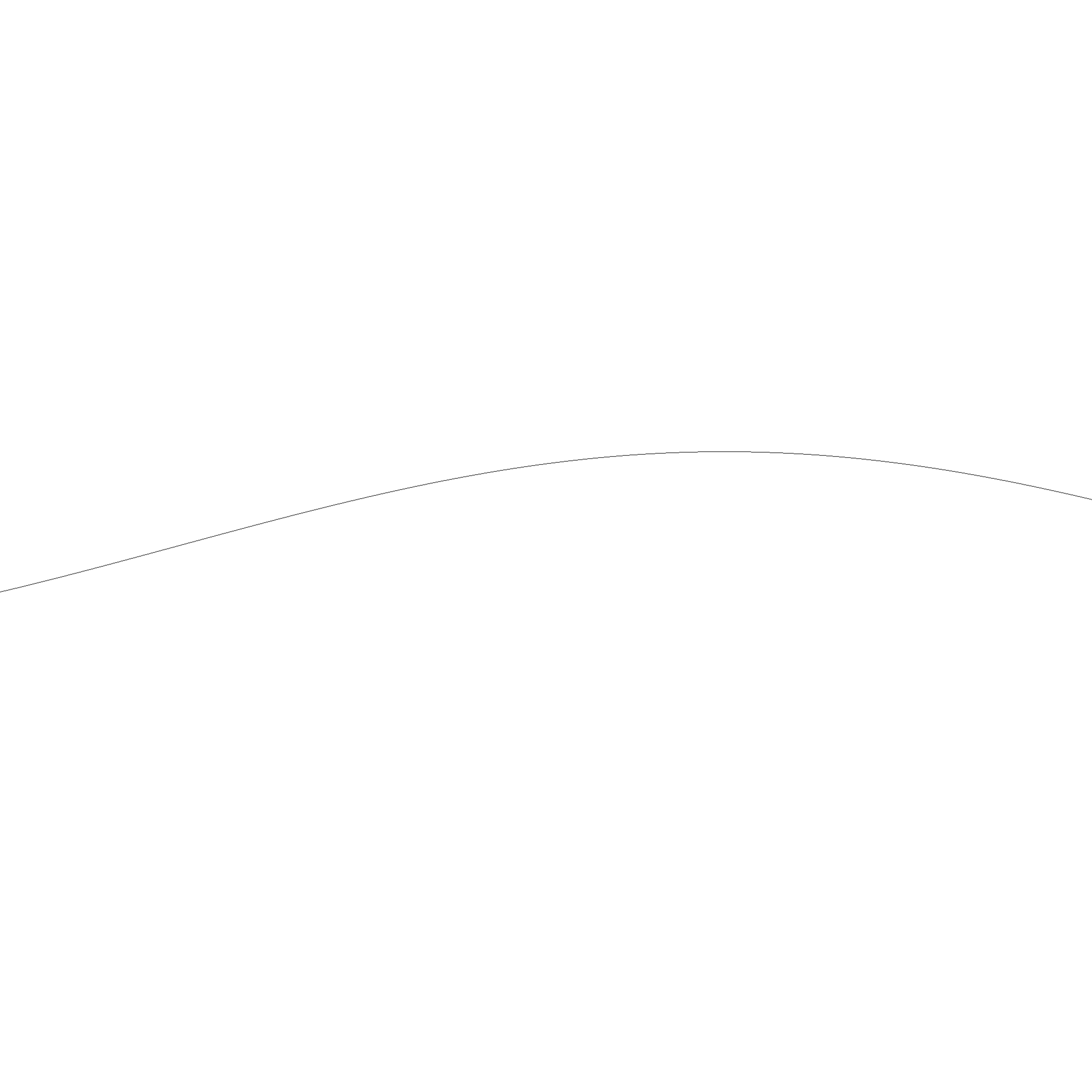} \label{fig:DisplacedJetVertex}}
  \caption{Hough transform algorithm applied to a displaced jet. Left: Simulated tracks originating from a
    displaced jet. Center: In the parameter space, the maxima are on a single sinusoid. Right: The
    second Hough transform identifies the sinusoid corresponding to the jet vertex.
    \label{fig:DisplacedJet}}
\end{center}
\end{figure}

\begin{figure}[!Hhtb]
\begin{center}
	\includegraphics[width=0.35\textwidth]{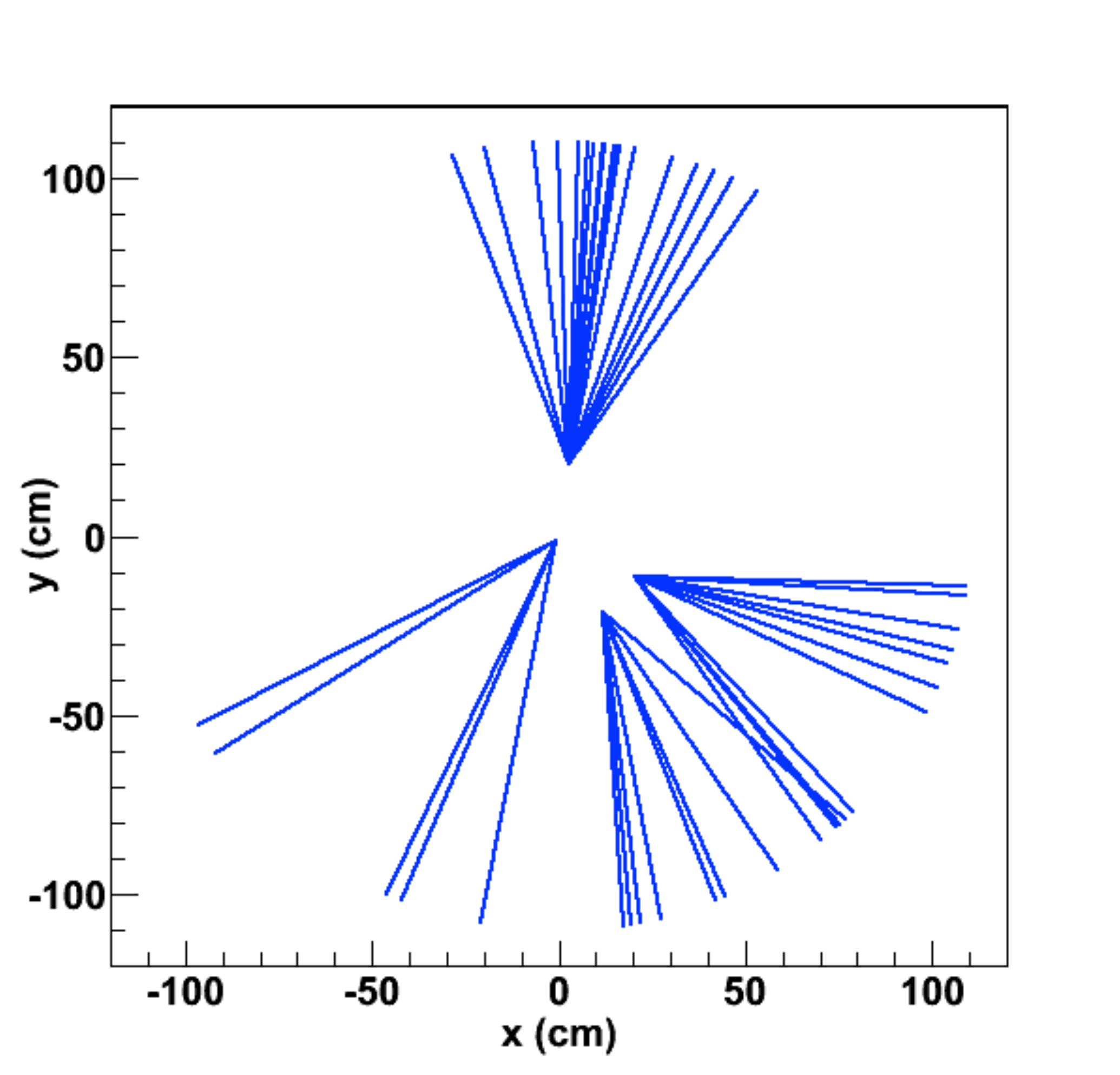}
	\includegraphics[width=0.30\textwidth]{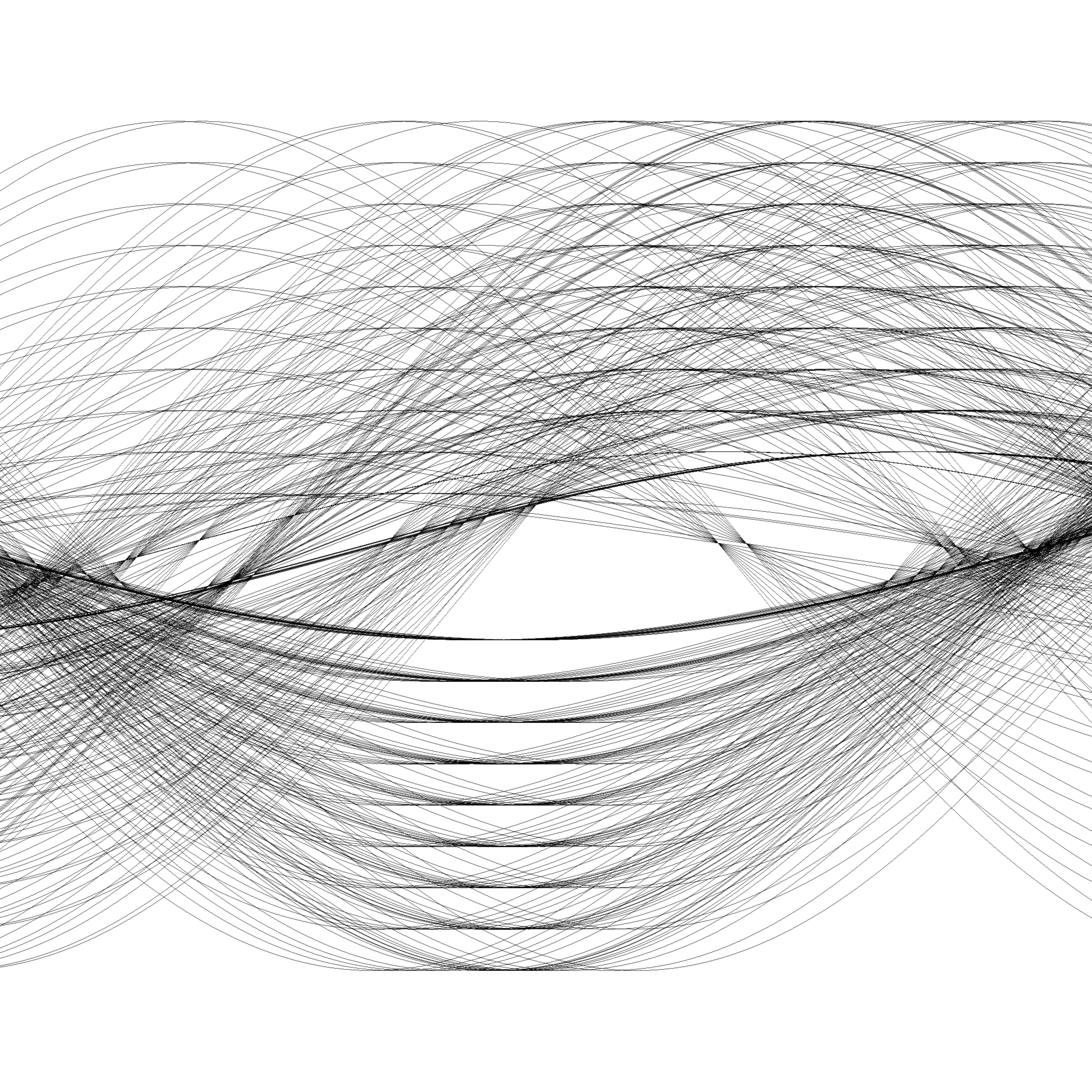}
	\includegraphics[width=0.30\textwidth]{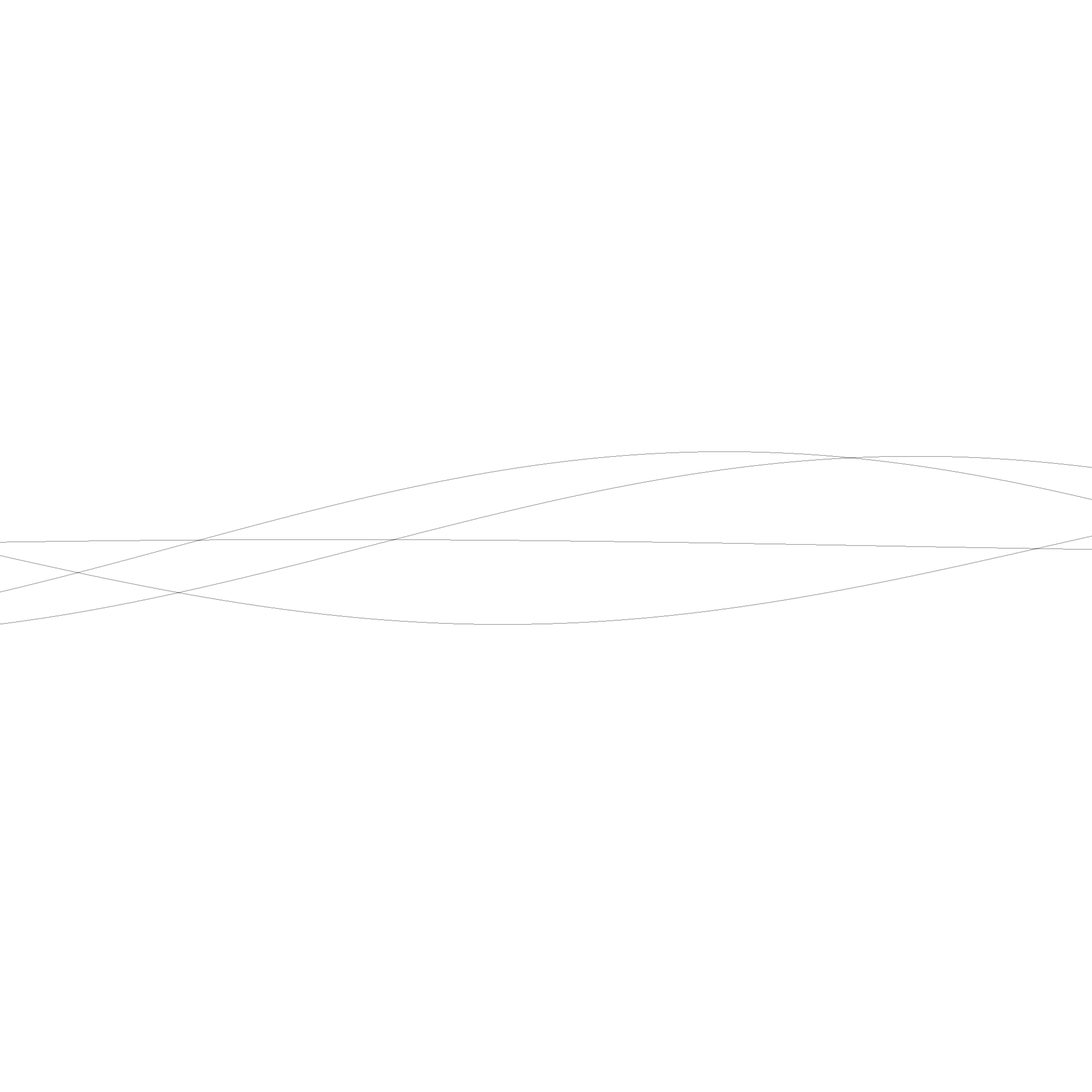}
	\caption{Hough transform algorithm applied to an event with multiple displaced jets. Left: Simulated
	tracks in the event. Center: The parameter space is difficult to visually interpret. Right: The
	second Hough transform identifies the sinusoids corresponding to the jet vertices.
	\label{fig:DisplacedJets}}
\end{center}
\end{figure}

The ability to handle displaced vertices provides the opportunity to identify interesting new physics such as non-prompt
jets and black holes. As an example a simulated set of tracks making up a jet, as shown in
Figure~\ref{fig:DisplacedJetTracks}, are considered.  The Hough transform is applied to the corresponding hits making up these
tracks to generate the parameter space displayed in Figure~\ref{fig:DisplacedJetAccumulator}. Recall that the parameter
space identifies candidate tracks as locations where many sinusoids cross one another. In
Figure~\ref{fig:DisplacedJetAccumulator} the locations of these crossings are visually apparent as the dark spots in
the parameter space.  However, the analysis is taken one step further by noting that the crossings all lie along a
single sinusoidal curve in the parameter space.  This single sinusoidal curve has been isolated in
Figure~\ref{fig:DisplacedJetVertex} and is significant because that one sinusoidal curve in the parameter space
corresponds to a point in the original x-y space.  More importantly, that point in the original x-y space corresponds to
the vertex of the displaced jet.  Interestingly, the most convenient way to implement this process is to use the Hough
transform a second time, this time applying it not to the hits in the x-y space but the crossings in the parameter
space.  Using the appropriate parameterization to look for sinusoidal curves instead of straight lines, the curve in
Figure~\ref{fig:DisplacedJetVertex} is identified and the vertex of the jet is now located.
Figure~\ref{fig:DisplacedJets} illustrates the same technique applied to an event with multiple displaced jets.

As previously mentioned it is the number of parameters, and not the details of the particular parameterization, that determines the computational cost of the Hough transform.  Therefore, compared to the performance results for a single Hough transform in Figure~\ref{fig:TimePerformance} the jet finding in Figure~\ref{fig:DisplacedJet} and Figure~\ref{fig:DisplacedJets} has double the computational cost because of the need to perform two transforms.
\section{Summary}

The quest for rare new physics phenomena has motivated the authors to propose a GPU-based enhancement of the trigger
system, made possible by the flexibility of the current trigger systems of the general-purpose detectors at the LHC. The
Hough transform algorithm, originally designed for machine vision problems, can be used as a tracking algorithm to
provide a different, complementary technique, which is a more holistic picture of the image as a whole. The algorithm
not only permits simultaneously finding any combination of prompt and non-prompt tracks, but also the development of new trigger
algorithms for exotic phenomena such as highly-displaced jets or black holes. These new topologies would represent a
``smoking gun'' for new physics, and so extending the trigger algorithm to be able to efficiently recognize them would
dramatically increase the reach of physics at the LHC. The proposed tracking algorithm would not replace the existing
combinatoric track finder, but will enhance its capabilities at the trigger level. The results presented here are
preliminary results meant to demonstrate the feasibility of the algorithm and its potential impact. Although it is
feasible to complete the testing of the new triggers proposed by the end of the shutdown, extensive work must be done to
improve tracking efficiency and purity for the general full tracking algorithm. In addition,
%Possible sentences for future work to better optimize CPU performance:
to better understand the relative performance of CPUs and GPUs, vectorization on the CPU to fully
utilize the Advanced Vector Extensions (AVX) capability present on modern CPUs from Intel and AMD will be implemented.  Although the GPU
might still show a compelling performance advantage for this class of computations, the 256 bit AVX capabilities should
increase the CPU performance and allow better quantification of that advantage.

%%%%%%%%%%%%%%%%
% References
%%%%%%%%%%%%%%%%

\end{document}